\documentclass[Journal]{IEEEtran}

\IEEEoverridecommandlockouts

\usepackage{graphics} 
\usepackage{graphicx}
\usepackage{epsfig} 
\usepackage{mathptmx} 
\usepackage{times} 
\usepackage{amsmath} 
\usepackage{amssymb}  
\usepackage{amsfonts}
\usepackage{float}
\usepackage{cite}
\usepackage{subfigure}
\usepackage[latin9]{inputenc}
\usepackage{booktabs}
\usepackage{graphicx}
\usepackage{algorithm}
\usepackage{algorithmicx}
\usepackage{algpseudocode}
\usepackage{setspace}
\usepackage{comment}
\usepackage{color}
\usepackage{url}
\usepackage{multirow}
\usepackage{threeparttable}

\usepackage{booktabs}
\usepackage{threeparttable}
\usepackage{multirow}

\newcommand{\tabincell}[2]{\begin{tabular}{@{}#1@{}}#2\end{tabular}}

\usepackage{verbatim}

\usepackage{colortbl}
\definecolor{mygray}{gray}{.8}
\definecolor{mypink}{rgb}{.99,.91,.95}
\definecolor{mycyan}{cmyk}{.3,0,0,0}

\makeatletter

\newcommand{\Rmnum}[1]{\expandafter\@slowromancap\romannumeral #1@}

\makeatother

\begin{document}

\title{AlphaZero Based Post-Storm Repair Crew Dispatch for Distribution Grid Restoration}

\author{Hang Shuai,~\IEEEmembership{Member, ~IEEE},~and Fangxing (Fran) Li,~\IEEEmembership{Fellow,~IEEE}

\thanks{H. Shuai, and F. Li are with the Department of Electrical Engineering and Computer Science, University of Tennessee, Knoxville, TN 37996 USA. \textit{Corresponding author: Fangxing (Fran) Li.}}

}

\date{}

\maketitle
\begin{abstract}
Natural disasters such as storms usually bring significant damages to distribution grids. 
This paper investigates the optimal routing of utility vehicles to restore outages in the distribution grid as fast as possible after a storm. 
First, the post-storm repair crew dispatch task with multiple utility vehicles is formulated as a sequential stochastic optimization problem.
In the formulated optimization model, the belief state of the power grid is updated according to the phone calls from customers and the information collected by utility vehicles. 
Second, an AlphaZero \cite{silver2017mastering} based utility vehicle routing (AlphaZero-UVR) approach is developed to achieve the real-time dispatching of the repair crews.
The proposed AlphaZero-UVR approach combines deep neural networks with stochastic Monte-Carlo tree search (MCTS) to give a lookahead search decisions, which can learn to navigate repair crews without human guidance.   
Simulation results show that the proposed approach can efficiently navigate crews to repair all outages.

\end{abstract}


\section*{Nomenclature}
\begin{IEEEdescription}[\IEEEusemathlabelsep\IEEEsetlabelwidth{$V_1,V_2,V_3,V_4$}]
\setlength{\parskip}{1pt}

\vspace{0.2cm}
\item[$\textbf{Variables}$]
\item[$A$] Trajectory of the utility vehicle.
\item[$H, G$] Binary indicator of received phone calls and location of the vehicle.
\item[$L$] Possible realizations of faulted power lines.
\item[$n_k^e$] Number of customers connected to node $k$ on circuit $e$.
\item[$P^{L}, P^{post}$] Prior and post probability of power line fault.
\item[$Q$] Mean action-value.
\item[$T_t^{travel}, T_t^{repair}$] Travel and repair time, respectively.
\item[$r,a$] Reward and decision (action) variable.
\item[$S$] State variable.
\item[$W$] Total action-value.
\item[$z, v$] Target value and computed value by neural network.
\item[$\theta$] Weights of neural network.

\vspace{0.2cm}
\item[$\textbf{Sets}$]
\item[$\ell ^ e$] Set of power line fault combinations on circuit $e$.
\item[$E$] Node set of road network.
\item[$B$] Node set of distribution grid.
\item[$S^{e}$] Set of segments on circuit $e$.
\item[$\Xi$] Circuits set.

\vspace{0.2cm}
\item[$\textbf{Superscripts and subscripts}$]
\item[$b, i$] Index of node and power line, respectively.
\item[$e$] Circuit.
\item[$l, t$] Hypothetical time-step and time index, respectively.
\item[$s$] Index of segment (power lines that trigger the same protective device).

\end{IEEEdescription}

\section{Introduction} \label{Introduction}
Climate change and global warming have caused an increase in both the frequency and intensity of extreme weather events, such as storms, floods, etc.
As the utility grids expose to natural environment and are quite vulnerable, natural disasters can significantly damage power distribution systems, which will lead to different levels of power outages.
Unfortunately, according to the study in \cite{kenward2014blackout}, extreme weather events (storms, droughts, and floods, etc.) have become the leading reasons for the U.S. electrical grid power outages.
More importantly, power outages caused by extreme weather events severely affected the normal operation of society and has brought huge economic losses. 
For example, the power outage costs of the February 2021 Texas winter storm, which left millions of people without power, are estimated to be \$80 billion - \$130 billion.

Currently, distribution grids, which directly supply electricity to commercial and residential customers, still remain vulnerable to extreme weather events.
To decrease the influence of extreme weather events on distribution grids, utilities have identified that measures should be taken to enhance the resilience \cite{house2013presidential} of distribution grids.
Regarding power system resiliency enhancement, researchers have conducted plenty of works recently and proposed many measures which can be divided into prior-to-events measures, during-events measures, and after-events measures, according to events unfolding process.
Prior to events, damage estimation modeling \cite{guikema2014predicting} and outage preventive planning strategies \cite{yuan2016robust,amirioun2017resilience,qiu2015optimal} (such as crews allocation) can largely help system operators to decrease the affections of weather events on distribution systems. 
During extreme weather events, topology switching strategy, generation re-dispatching and load shedding strategies \cite{yan2018coordinated, trakas2017optimal} are proposed to increase the resiliency of distribution systems.
Finally, after extreme weather events passed, utility will assess the actual damages, then recover power supply through system restoration and load restoration strategies \cite{hou2011computation, sun2010optimal, arif2018optimizing} by coordinating all the generation resources and repair crews.

In this paper, we investigate the restoration strategy of distribution systems after storm events.
More specifically, this work focuses on the utility vehicle routing (UVR) problem with the goal of dispatching repair crews to fix all faults in the system as fast as possible.
After the power outage caused by storms reported, the utilities need to estimate possible locations of damages firstly by using outage and fault location methods, then repair crews will be scheduled to fix all fault equipment.
Considering repair crews usually driving utility vehicles to fix outages in systems, so repair crew dispatch problem is largely a UVR problem.

Regarding the post events utility vehicle routing, there exists some researches.
For instance, a constraint injection based optimization algorithm was proposed in \cite{van2011vehicle} to solve the UVR problem, with the assumption that system operators can get the precise locations of all faults according to the engineer's operational knowledge.
Reference \cite{7812566} proposed a two-stage outage management method for distribution system repair and restoration problem (DSRRP).
In the first stage, the repair tasks are clustered to crews according to the damage location and damaged components information.
In \cite{9115714}, a multiperiod distribution system restoration model, which is in response to multiple outages caused by natural disasters, was proposed.
A resilient after disaster recovery scheme was proposed in \cite{8642442} to co-optimize distribution system restoration with the dispatch of repair crews and mobile power sources.
Similar with \cite{van2011vehicle}, the works in \cite{7812566,9115714,8642442} assume that system operators can obtain the precise locations of all faults in the grid.
However, according to \cite{chen2017modernizing,7444207}, this assumption is too idealistic.
Although researchers have developed a variety of  outage area and fault location methods by using different input data (e.g. non-electrical data, electrical data, network data, and measurements), precisely detect faults and locations in distribution systems is still a tough task (see \cite{1007917,956755,executive, bahmanyar2017comparison}).
For instance, traditional distribution grids identify power outages through trouble calls from customers due to the weak situation awareness ability of utilities \cite{chen2017modernizing}.
Thus, some utilities may not aware of outages and possible locations of faults until receiving the outage-reporting calls from customers \cite{chen2017modernizing,7444207,executive}.
With the development of modern distribution networks, advanced metering infrastructure (AMI) enables the utilities to remotely read consumer consumption records, and provides a new source of information for outage location.
However, the overall procedure of the AMI-based outage location is similar to the trouble call-based methods mentioned above, which usually estimates the most likely area or location of faults\cite{bahmanyar2017comparison}.
The benefit of the AMI-based approaches is that the operators do not need to wait for a sufficient number of customer calls to locate the outage areas.
In this way, developing a post storm restoration strategy which does not rely on the precise location information of faults is meaningful for utilities.

Different from the above research works, this paper focuses on the post-storm UVR problem without knowing the precise locations of damages.
Considering the damage status of distribution systems after a storm is usually unknown, reference \cite{al2016information} proposed an information collecting vehicle routing model for the first time, which uses trouble calls from customers and fault information dynamically collected by crews on a vehicle to create beliefs about outages.
More importantly, in reference \cite{al2016information}, the UVR problem was formulated as a sequential stochastic optimization model and a Monte Carlo tree search (MCTS) based vehicle navigation strategy was innovatively proposed.
The work of \cite{al2016information} paves the way for the application of machine learning methods to distribution grid restoration.
Based on the work of \cite{al2016information}, the authors of this paper proposed an Open Loop Upper Confidence Bound for Trees (OLUCT) algorithm based utility vehicle routing strategy in \cite{hang2020IEEEPES}.
However, the traditional tree search methods adopted in \cite{al2016information} and \cite{hang2020IEEEPES} need a huge number of iterations, which is relatively time consuming, to find the most efficient path.
Besides, reference \cite{al2016information} and \cite{hang2020IEEEPES} simplified the post-storm UVR problem by assuming that there is only one utility vehicle in the system.
Thus, the discussion in \cite{al2016information} and \cite{hang2020IEEEPES} are all focused on single UVR problem.
Nevertheless, utilities usually have multiple repair vehicles which are standby after a storm.
It is more realistic to develop a post-storm repair crew dispatch model with multiple repair vehicles and design a multi-vehicle routing algorithm.
Solving the multi-vehicle routing problem without knowing the precise location of faults is a very challenging task.
Thus, we will use the advanced deep reinforcement learning (DRL) technique to solve the optimization problem.

Recently, with the development of DRL, plenty of machine learning algorithms have been proposed and obtained superhuman performance in a variety of sequential decision problems.
For instance, AlphaGo \cite{silver2016mastering} and AlphaZero \cite{silver2017mastering} algorithms achieved superhuman performance in board games including the game of Go.
Specially, AlphaZero convincingly defeated world-champion players without human guidance and domain knowledge beyond game rules, by tabula rasa reinforcement learning from data generated using self-play mechanism.
The core of AlphaZero algorithm \cite{silver2017mastering} is the combination of MCTS with a deep neural network (DNN) which consists of a policy network and a value network.
Once well-trained off-line, the policy network and value network can effectively guide the tree search process to make optimal actions.
The authors of this paper have discussed several potential AlphaGo-like application scenarios in power systems in \cite{FranLi2018}, and also investigated the application of MuZero \cite{schrittwieser2020mastering} in microgrid optimal scheduling problem in \cite{Hang2021TSG}.
The research works in \cite{FranLi2018,Hang2021TSG} indicate that AlphaGo/AlphaZero/MuZero based algorithms are promising for solving challenging problems in power systems.

To this end, this paper focuses on post-storm repair crew dispatch problem with multiple utility vehicles and investigates the application of AlphaZero for post-storm UVR to restore distribution grids as fast as possible.
With the advantages of AlphaZero approach, we demonstrated that the proposed navigation strategy can make real-time crew dispatch decisions according to the current system state, which is critical for this specific application problem.
The main contributions of this work are summarized as follows:
\begin{itemize}
\item[1)] A post-storm repair crew dispatch model with multiple utility vehicles is formulated.
\item[2)] An AlphaZero \cite{silver2017mastering} based post-storm utility vehicle routing (AlphaZero-UVR) algorithm is proposed to navigate the crews to restore the distribution grid as fast as possible.
\item[3)] AlphaZero algorithm has been applied to playing board games which is with a deterministic state transition function.
However, the state transition of the multiple time-step optimization problem in this work is stochastic as the actual damage status of the unvisited lines are unknown. 
To apply AlphaZero to this problem, we modified the original AlphaZero algorithm by combining DNN with stochastic MCTS method.
\item[4)] The simulation results on a 8-node test system and a modified IEEE 123-node distribution system demonstrate the effectiveness of the proposed crew dispatch strategy.
\end{itemize}

This paper is organized as follows. 
The stochastic UVR problem is formulated as a Markov Decision Process (MDP) in Section II. 
Section III presents the developed AlphaZero based utility vehicle routing algorithm. 
The simulation results on two distribution systems are given in Section IV. 
Section V concludes the paper.

\section{Post-storm Repair Crew Dispatch Problem}
The repair crew dispatch problem is modeled as the UVR problem.
In this section, the distribution network fault location method used in this work is presented firstly.
Then, the trouble call-based UVR problem is introduced. 
Finally, the post-storm UVR problem with multiple utility vehicles is formulated as a Markov Decision Process (MDP).
Note that, in \cite{hang2020IEEEPES, al2015probability}, a trouble call-based UVR model with single utility vehicle was formulated.
In this work, based on the previous work, we extend the UVR model for multiple utility vehicles, and a new AlphaZero based optimization method is designed to solve the problem in this paper. 
\subsection{Post-storm fault location method}
\begin{figure}[t]\centering
\includegraphics[width=3.0in]{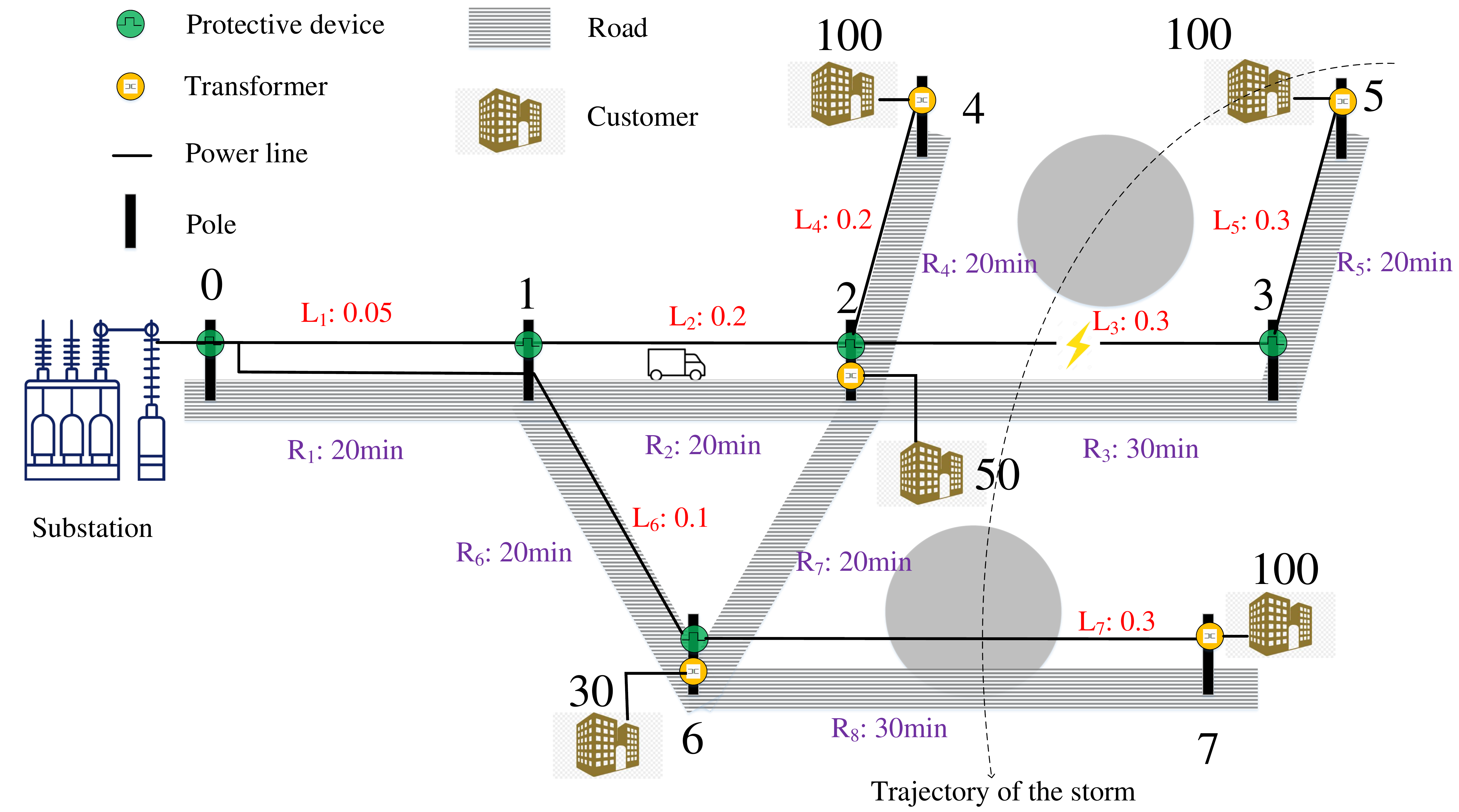}
\caption{Utility vehicle routing in a distribution system \cite{hang2020IEEEPES}.} \label{fig:DistributionPowerSystem}
\vspace{-1em}
\end{figure}

The distribution system includes substations, overhead power lines, poles, transformers, protective devices, and customers, as shown in Fig. \ref{fig:DistributionPowerSystem}.
The electricity is distributed through power lines and delivered to customers.
Customers are connected to transformers which are fixed on the poles. 
Besides, to isolate faults, a number of protective devices, such as protective relays, disconnect switches, etc., are also installed on the poles. 
Once a protective device is triggered, all the downstream customers will suffer from power failure. 
For instance, when the protective device on pole 2 in Fig. \ref{fig:DistributionPowerSystem} is triggered, then, the downstream customers connected to node 4 and 5 will lost power supply.

The input data used by fault location algorithms can be classified into four groups \cite{bahmanyar2017comparison}, namely non-electrical data (e.g., customer calls, weather data), electrical data (e.g., SCADA data, smart meter data), network data (e.g., line, network topology), and measurements (e.g., substation voltage and current).
Based on the required inputs, the outage location methods can be classified to trouble call-based methods, historical data-based methods, fault indicators based algorithms, AMI-based methods, and algorithms using a combination of different sources of data.
In this work, we adopt the trouble call-based method to estimate the possible locations of faults after storms.
So, the utility identifies the damages in the system according to the phone-call reports received by electric utility center (EUC), weather data, and other network data.

After a storm passed, all the equipment in the system has a fault possibility.
This possibility is determined by many factors, such as the trajectory of a storm, the strength and the duration time of a storm weather, and the anti-storm capacity of a device, etc.
So, after storms passed through, EUC will evaluate the damage probabilities of all the devices, then dispatch crews to repair all damages.
To evaluate the damage status, the information that the EUC can take use of including the trouble calls from customers, the actual damage status information collected by the crews on their passed paths, and the prior fault probabilities of devices which can be evaluated according to the storm forecast information and the operation experiences from the system operators.

Based on references \cite{al2015probability, al2016information, hang2020IEEEPES}, the posterior probability of power line $i$ on circuit $e$ being in fault  at time $t$ given the phone calls $H_t$ and the trajectory of the vehicles $A_{t-1}$, can be calculated using Bayes' theorem as follows:
\begin{equation}\label{EQ1}
\begin{aligned}
p(L_{t, i}^e = 1|H_t, A_{t-1}) = \frac{\Sigma_{{L_t^e \in \{{\ell}^e}\}_{L_{t, i}^e = 1}} p(H_t|L_t^e) p(L_t^e | A_{t-1}, O_{t-1})}{\Sigma_{{L_t^e \in \{{\ell}^e}\}} p(H_t|L_t^e) p(L_t^e | A_{t-1}, O_{t-1})}
 \end{aligned}
\end{equation}
where $A_t$ represents the vehicle's trajectory up to time $t$.
In other words, $A_t$ consists of all the past routing decisions.
$H_t = \{H_{t, b}^e: b \in B\}$ is a possible realization of the received trouble calls.
$H_{t,b}^e$ represents whether the EUC received reporting calls from node $b$ on circuit $e$ by time $t$.
For a circuit $e$, the possible realizations of the faulted power lines is represented by the vector $L_{t}^e$, and the $i$th element of the vector is $L_{t,i}^e$.
When power line $i$ is faulted at time $t$, $L_{t, i}^e$ equals to 1.
So, $\{\ell ^ e\}_{L_{t, i}^e = 1}$ is a subset of vectors of $L^e$ where power line $i$ is faulted.
The likelihood $p(L_t^e | A_{t-1})$ is the prior fault probability of power lines.
The prior fault probability can be obtained according to the operational experience and the storm information.
Besides, prior fault probability keeps being updated in the following time periods using new observations of the system.
$p(H_t|L_t^e)$ is the likelihood of the calls given the power line faults on circuit $e$, which can be calculated by Eq. (\ref{EQ2}).
\begin{equation}\label{EQ2}
p(H_t|L_t^e) =
\left\{
\begin{aligned}
&\prod_{N_{i} \in \Psi_{J(L_t^e)}} 1 - (1 - \rho_i)^{n_i}, \, if \; L_t^e \in Z(H_t) \\
&0, otherwise
 \end{aligned}
 \right.
\end{equation}
where $Z(H_t)$ represents all the combinations of power lines that will cause all customers in $H_t$ to call if faulted.
$\Psi_J(L_t^e)$ denotes the set of affected nodes (lose power) if the protective devices of the faulted power lines $L_t^e$ are triggered.
$\rho_i$ is the customer calling probability when they suffer from power outage.
$n_i$ is the total number of customers connected to node $i$.

As indicated by the research in \cite{al2015probability}, the computational complexity of the fault probability model (\ref{EQ1}) is mainly affected by the number of power lines in a power grid.
Fortunately, there are several approaches that can largely decrease the computational complexity such as power lines aggregation and Monte Carlo simulation.
Readers are referred to reference \cite{al2015probability} for more details.

\subsection{Post-storm UVR problem}
\begin{figure}[t]\centering
\includegraphics[width=2.8in]{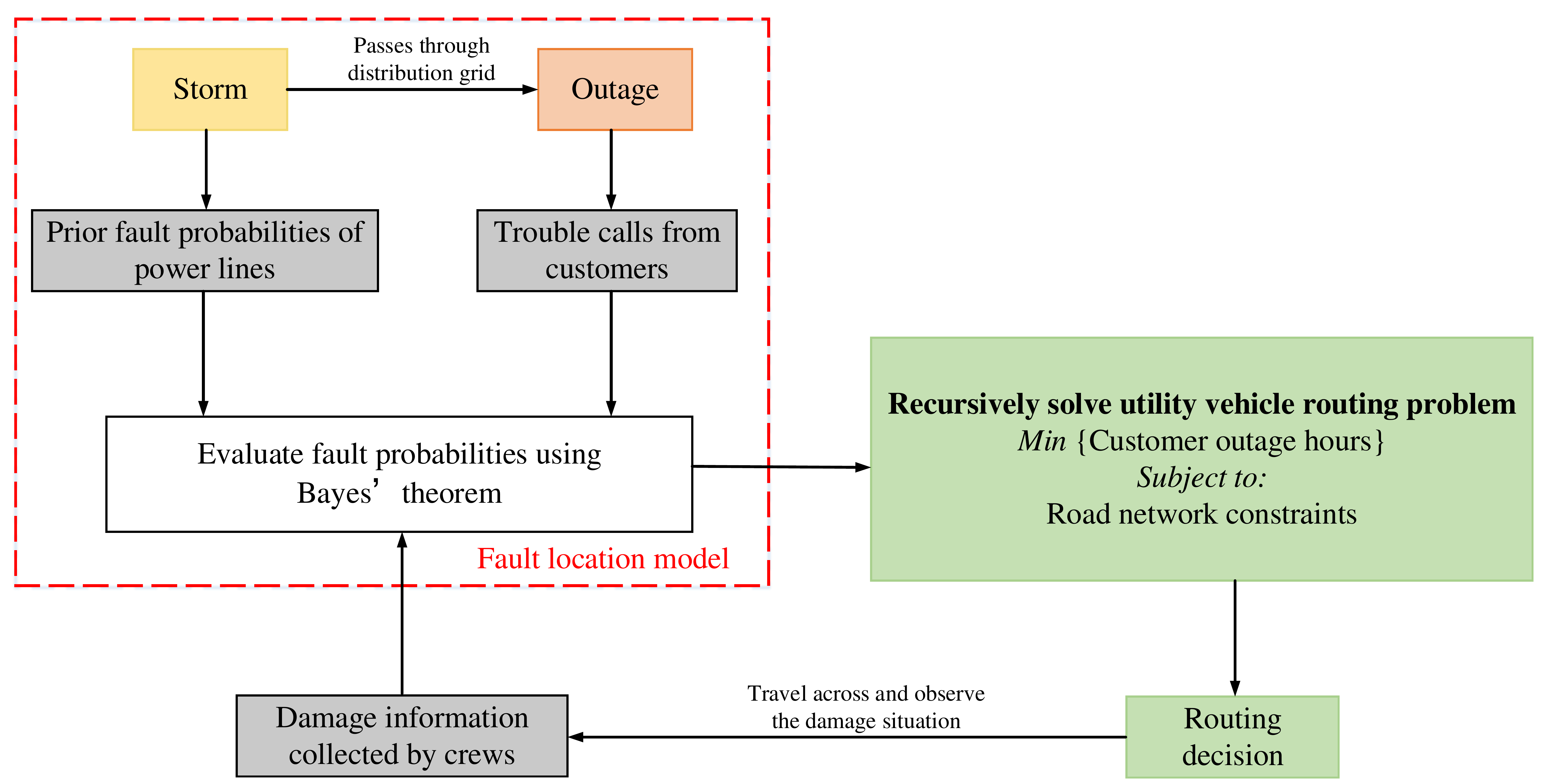}
\caption{Illustration of post-storm UVR problem.} \label{fig:UVR-Sch}
\vspace{-1em}
\end{figure}

After evaluated the fault probabilities of all devices according to the reporting calls from customers and the latest damage information collected by utility vehicles, EUC needs to make real-time routing decisions recursively. 
The real-time decision is obtained by solving the utility vehicle routing (UVR) problem that minimizes the outage-hours of the system.
In addition, the optimization problem must subjects to road network constraints.
The road network constraints mainly determine the feasible action space of the vehicle.
For instance, if the vehicle located in node 2 of the system shown in Fig. \ref{fig:DistributionPowerSystem}, the feasible destination nodes of the vehicle at the next time-step will be 1, 3, 4, and 6.
The proposed UVR scheme can be summarized as Fig. \ref{fig:UVR-Sch}.
Although the utility vehicle routing optimization scheme shown in Fig. \ref{fig:UVR-Sch} adopts the trouble call-based outage location method, other outage/fault location methods (e.g., AMI based methods) can be easily applied in the architecture, as the UVR problem only needs to know the fault probabilities of devices output by the fault location model.
The power flow constraints are not included in the UVR optimization problem, as the damage status of power lines that unvisited by the vehicles is unknown.

\subsection{Optimization model of the post-storm UVR problem}
After the storm passed, EUC will dispatch multiple utility vehicles (crews) travel across the system to fix all possible damages, with the objective of scheduling the crews to restore the grid as quickly as possible.
Based on the work of \cite{al2016information, hang2020IEEEPES}, we formulate the sequential stochastic optimization problem as a Markov Decision Process (MDP).
However, different from references \cite{al2016information, hang2020IEEEPES} which focus on single vehicle routing, this work investigates the multiple utility vehicles routing problem.

\begin{figure}[t]\centering
\includegraphics[width=2.8in]{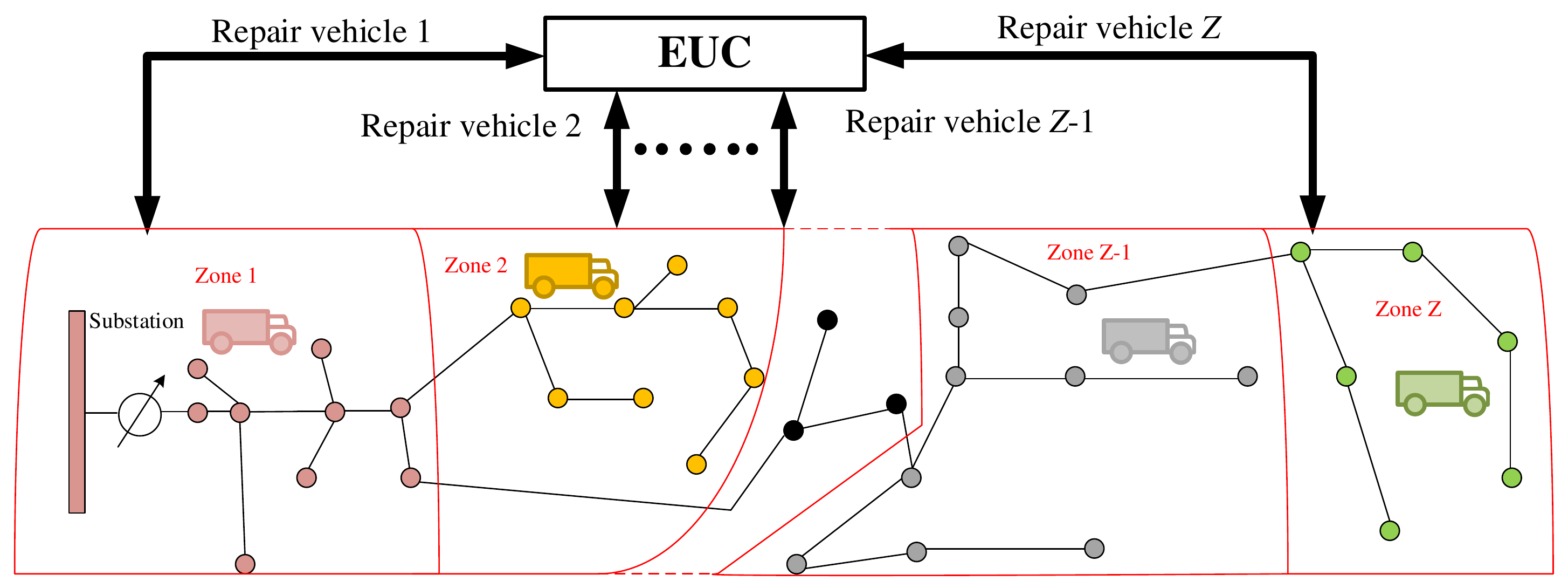}
\caption{Illustration of the multi-vehicle routing in a distribution system. The red line represents boundaries of each zone.} \label{fig:Multi-vehicle}
\vspace{-1em}
\end{figure}

The designed post-storm multi-vehicle scheduling architecture is shown in Fig. \ref{fig:Multi-vehicle}.
The distribution grid is divided into $Z$ zones, and each repair vehicle travels across a specific zone to fix all the damages found in that area.
When the vehicle travels from the starting node to the end node of a power line, it will observe the damage status of the line, and the crews on the vehicle will report the observation to the EUC.
Then, the EUC will update the fault probabilities of all power lines in the whole distribution system using Eq. (1).
Once a vehicle reached the end node of a line and didn't find any fault or it found a fault and repaired the damage, the vehicle will send a scheduling request to the EUC in order to get the optimal action of the next time-step. 
After the EUC received the request, it will calculate the optimal travel action of this vehicle according to the updated fault probabilities of all power lines and the feasible action space of the vehicle.
The feasible action space is related to the current position of the vehicle and the topology of the road system.   
In this work, the EUC will use the well-trained AlphaZero agent to get the optimal action of the vehicle.
After received the action signal from the EUC, the vehicle will go to the destination node. 
The above process is repeated until all the possible damages have been repaired.
If the EUC received scheduling requests from multiple vehicles at the same time, each time the EUC will select one repair vehicle from the request queue according to the priority to give scheduling decision until the queue becomes empty. 

In the following section, the basic elements of the MDP model will be defined.
\subsubsection{State variable}
\begin{equation}\label{EQ3}
\begin{aligned}
S_t = \Big\{G_t, P_t^L, H_t\Big\} = \Big\{G_t, P_t^{L,post}\Big\}
 \end{aligned}
\end{equation}
The state variables consist of three parts, namely, the physical state $G_t$, the belief state $P_t^L$, and the informational states $H_t$. 
$G_t$ is location of the vehicle under scheduling at current time-step, and $P_t^L$ is the prior fault probabilities of power lines.
The elements of the vector $P_t^L$ are $p(L_{ti}^e = 1|A_{t-1})$.
Using $P_t^L$ and $H_t$, we can get the posterior fault probability $P_t^{L,post}$ according to Eq. (\ref{EQ1}).
\subsubsection{Decision variable}
\begin{equation}\label{EQ4}
\begin{aligned}
a_t^z = (a_{t, ij}^z)_{i,j \in E}
 \end{aligned}
\end{equation}
where $a_t^z$ is the routing decision of vehicle $z$ at time-step $t$.
If the $z$th vehicle goes from node $i$ to $j$ at time-step $t$, then $a_{t, ij}^z = 1$.
\subsubsection{Transition function}
\begin{equation}\label{EQ5}
\begin{aligned}
&G_{t+1} = j, \; if \; a_{t, ij}^z = 1 \; and \; I_{t+1}^z = 1
 \end{aligned}
\end{equation}

\begin{equation}\label{EQ6}
\begin{aligned}
p(L_{t+1, j}^e|\sum_i \sum_v a_{t, ij}^v = 1) = 0
 \end{aligned}
\end{equation}
\begin{equation}\label{EQ7}
\left\{
\begin{aligned}
&H_{t+1} = H_{t}, if \ \hat H_{t+1} = 0 \\
&H_{t+1} = 1, if \ \hat H_{t+1} = 1
 \end{aligned}
 \right.
\end{equation}
where $I_{t+1}^z = 1$ represents the $z$th vehicle sends the scheduling request at time-step $t+1$.
$\hat H_{t+1}$ represents the indicator of newly arrived trouble calls at time-step $t+1$.
In this work, once the faulted power lines were fixed by crews, the fixed lines won't be in fault again in the following times, as shown in Eq. (\ref{EQ6}).
It is worth noting that if the fault probability of the $j$th power line is updated, the fault probabilities of other power lines also need to be updated using Eq. (\ref{EQ1}).
\subsubsection{Objective function}
The objective function of the UVR problem is given by:
\begin{equation}\label{EQ8}
\begin{aligned}
F &= \min_{\pi} \verb"E" \Bigg\{ \sum_{t=0}^T C_t (S_t, A^{\pi}(S_t))|S_0 \Bigg\} \\
&= \max_{\pi} \verb"E" \Bigg\{ \sum_{t=0}^T r_t (S_t, A^{\pi}(S_t))|S_0 \Bigg\}
 \end{aligned}
\end{equation}
The optimization objective is to minimize the cumulative custom outage hours.
$\pi$ represents the policy adopted in the optimization.
$C_t(S_t,a_t)$ is the custom outage hour when the system at state $S_t$ and takes decision $a_t$.
$r_t(S_t,a_t)$ is the reward function and $r_t(S_t,a_t) = - C_t(S_t,a_t)$.
In Eq. (8) and the following equations, we abbreviate $a_t^z$ as $a_t$.

According to Bellman's optimality, the optimal policy can be solved by:
\begin{small}
\begin{equation}\label{EQ9}
\begin{aligned}
A_t^*(S_t) = arg \max_{{a_t} \in {\chi_t(S_t)}} \Bigg( r_t(S_t,a_t) + \max_{\pi} \verb"E" \Bigg\{
\sum_{\tau=t+1}^T r_{\tau} (S_{\tau}, A_{\tau}^{\pi}(S_{\tau}))|S_0 \Bigg\} \Bigg)
 \end{aligned}
\end{equation}
\end{small}
Note that $r_t(S_t,a_t)$ cannot be exactly calculated during the repairing process even though $S_t$ and $a_t$ are known.
Because the number of restored customers after the utility vehicle visited a power line depends on upstream and downstream outages of the system.
Unfortunately, these outages are uncertain.
However, the expected value of the reward can be evaluated as follows:
\begin{equation}\label{EQ10}
\left\{
\begin{aligned}
&r_t(S_t,a_t) = - \Bigg(\sum_{e \in \Xi} \sum_{s \in S^{e}} \big(1 - \Pi_{k \in K_s} p(L_{t, k}^e = 0) \big) \sum_{k \in s} n_k^e \Bigg) \cdot T_t \\
&T_t = T_t^{travel} + T_t^{repair}
 \end{aligned}
 \right.
\end{equation}
where $K_s$ is the set of lines (if faulted) that will cause the power supply failure of segment $s$.
$T_t$ is the time needed to go from the current node to the destination node, which includes the travel time and the repair time.
The travel time is determined by the length of the road and the driving speed of the vehicle.
The repair time is related to the fault locations.

From the above equations, it can be found that the UVR problem is formulated as a sequential stochastic optimization problem.
Solving the above problem faces several challenges. 
First, how many customers are restored after a line fixed by the crews is uncertain since the incomplete information of other faults. 
Second, the changing belief states (keep updating during the repairing process) of the grid further increases the difficulty of the problem solving.
Traditional mathematical optimization approaches (i.e., integer programming) are difficult to solve this optimization problem.
Reference \cite{al2016information} proposed a MCTS based single utility vehicle routing approach which is a machine learning based optimization algorithm.
Nevertheless, the performance of the MCTS based approach still has much room for improvement.
Reinforcement learning (RL) methods have been widely adopted to solve MDP problems and plenty of research works have demonstrated that DRL methods can effectively solve many challenging tasks, like game of Go \cite{silver2017mastering}.
So, we investigate using the state-of-the-art DRL technique developed by \cite{silver2017mastering}, to solve the multi-vehicle routing problem in this work.

\section{AlphaZero Based Post-storm Utility Vehicle Routing Algorithm}
In this section, an AlphaZero \cite{silver2017mastering} based post-storm UVR algorithm is proposed to guide the crews to repair grid outages as fast as possible.
The challenges of directly using AlphaZero algorithm to solve the UVR problem are presented.
In this way, a modified AlphaZero architecture is developed for the problem.
Then the AlphaZero based utility vehicle routing algorithm is designed.

\subsection{AlphaZero algorithm}
AlphaZero \cite{silver2017mastering} is a model-based deep reinforcement learning algorithm developed by DeepMind in 2018 to paly the games of chess, shogi, and Go, which has achieved superhuman performance.
Similar with AlphaGo, the key issue of AlphaZero is to integrate deep learning technique into Monte-Carlo tree search (MCTS).
The advantage of AlphaZero is that it does not need human guidance and domain knowledge of the game except the game rules, and learns to play the game entirely by self-play.
Besides, a single DNN architecture that contains policy outputs and value output is proposed in AlphaZero.

The general principle of AlphaZero is that it plays against itself to generate training data, with each side of the game choosing actions by MCTS strategy, and the generated self-play game data are sampled to continually \textit{train the deep neural network}.
Then, using the latest DNN, the new game data are generated by \textit{self-play}.
These procedures repeat until the algorithm converged.
The \textit{self-play} and \textit{neural network training} are conducted in parallel, each improving the other.
During the self-play procedure, MCTS is used to get the optimal action of each time-step, while the MCTS uses the neural network to guide its simulation as shown in Fig. \ref{fig:AlphaZeroTree} (a).
In the figure, every node is associated with a system state $S$, and the edge $(S, a)$ of the search tree contains the prior probability of selecting the edge $P(S, a)$, the visit count $N(S, a)$, the total action-value $W(S, a)$, and the mean action-value $Q(S, a)$.
The search tree shown in the figure is constructed by conducting a predefined number of simulations, and each simulation consists of the \textit{selection}, \textit{expansion}, and \textit{backpropagation} procedures.
Details about the MCTS simulation are elaborated in the following section.

\subsection{AlphaZero based post-storm utility vehicle routing approach}
Plenty of machine learning algorithms need massive training data.
However, extreme weather events are usually low probability, which means EUC only has limited event data and the corresponding action (crew dispatch) data.
Moreover, we are not sure whether the historical action data is optimal.
So, to solve the post-storm UVR problem by using machine learning approach, one challenge is that we do not have enough (labeled) data to train the designed algorithm.
Luckily, the self-play mechanism proposed in AlphaZero provides a good solution to this problem.
However, playing game of Go is very different from the UVR problem in this work.
Actually, to apply AlphaZero to the UVR problem, it faces challenges brought by the differences between board games and the problem in this work.
The original AlphaZero algorithm is designed for the two-player games, while the UVR problem can only be viewed as a single-player game.
Fortunately, this difference will not hinder the application of the algorithm in the UVR problem and it even simplifies the application of the algorithm as the agent does not need to change the players during the training and online application processes. 

Besides, the original AlphaZero algorithm is used to solve the MDP problems with a deterministic transition function.
For instance, in the game of Go, the next board state is deterministic after taking a particular action.
However, the transition function of the UVR problem in this work is stochastic.
The uncertainty comes from the partially observation characteristic of the post-storm distribution system.
For example, at time $t$, we assume the vehicle is located in node 1 of Fig. \ref{fig:DistributionPowerSystem}, and makes the decision (action) to head to node 2, which means it will pass through line $L_2$.
After the vehicle passed the power line, the actual damage status of $L_2$ will be observed, either damaged or not.
From Eq. (\ref{EQ1}), different observed damage status will lead to different evaluation value of the other lines' fault probabilities.
So, even if the vehicle takes the same navigation decision, it has the probability of transferring to different states.

\begin{figure}[t]\centering
\includegraphics[width=2.8in]{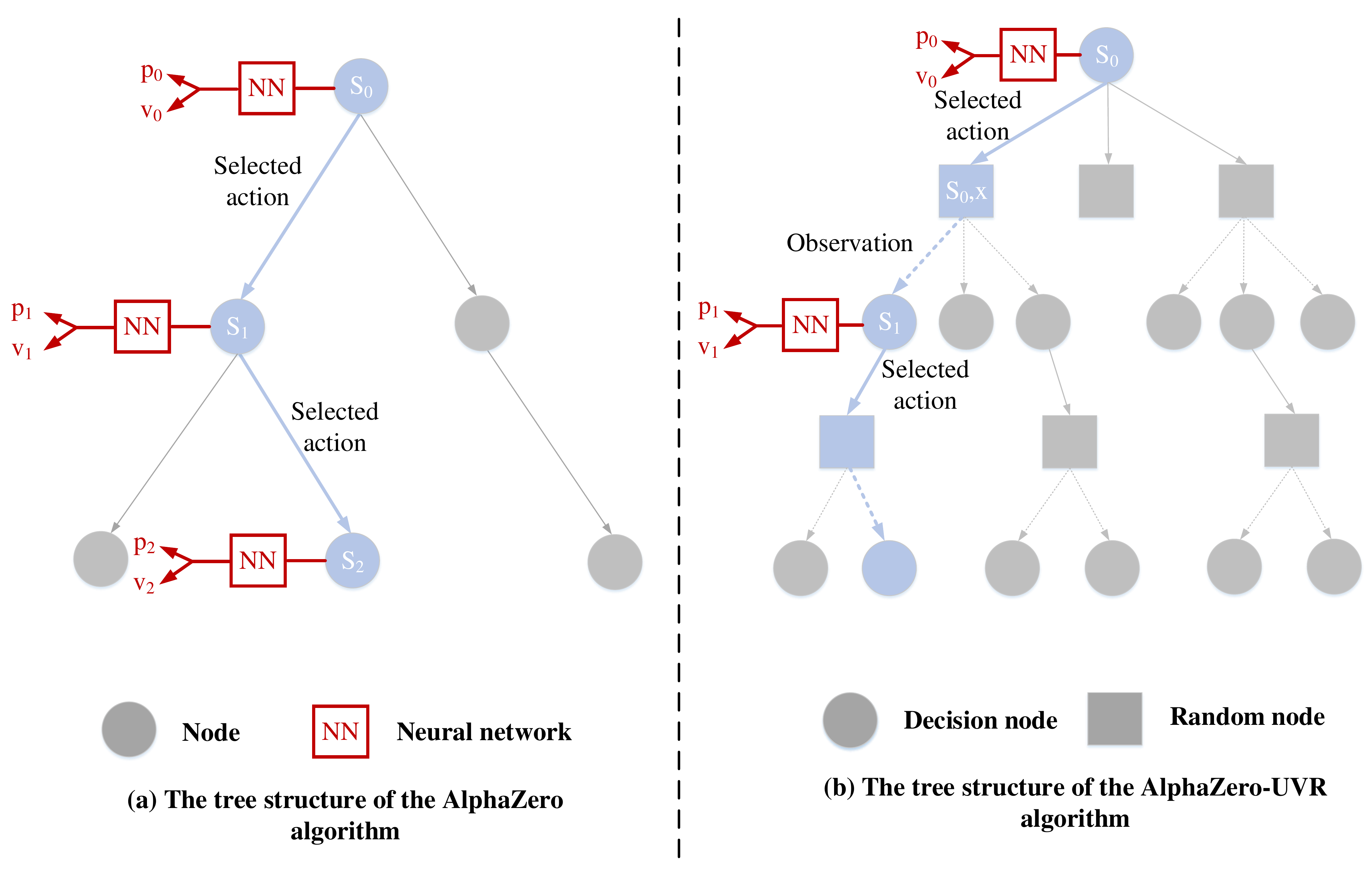}
\caption{Tree structure of the original AlphaZero algorithm \cite{silver2017mastering} and the developed AlphaZero-UVR algorithm.} \label{fig:AlphaZeroTree}
\vspace{-0.5em}
\end{figure}

\textit{1) \textbf{Combine stochastic MCTS with DNN}}:  To deal with the stochastic transitions in the UVR problem, we combine the stochastic MCTS method with DNN, and extend the original AlphaZero algorithm to deal with the stochastic dynamics in the environment (distribution grid).
The tree structure difference of the developed AlphaZero-UVR approach and the original AlphaZero algorithm is shown in Fig. \ref{fig:AlphaZeroTree}.
In the AlphaZero-UVR algorithm, there includes random nodes (contain state - action pair) and decision nodes (contain state information).
The state information of decision nodes is defined as Eq. (\ref{EQ3}).
The action space is defined in Eq. (\ref{EQ4}). 
Starting from the decision node, different random nodes are created by taking different actions (crew routing decisions).
The action $a$ is selected from the feasible action space which is determined by the current location of the vehicle and the road network.
The uncertainties in the distribution grid will lead to various observations $o$ (the power line passed through is damaged or not), then it results in different decision nodes.
For each action $a$ from decision node there is an edge that stores the statistics $\{ N(S, a), W(S, a), Q(S, a), P(S, a), r(S, a)\}$, and for each observation $o$ from random node there is an edge that stores the statistics $\{ N(S, a, o)\}$.
Note that $r(S, a)$ can be calculated by Eq. (\ref{EQ10}).

\textit{2) \textbf{MCTS simulation in AlphaZero-UVR}}: Since the tree structure differences, the MCTS simulation, which includes \textit{selection}, \textit{extension}, and \textit{backpropagation}, of the AlphaZero-UVR algorithm is different from the original AlphaZero algorithm.
The authors designed the MCTS simulation procedures for the UVR problem, as as shown in Fig. \ref{fig:MCTS in AlphaZero-UVR} and below:
\begin{figure}[t]\centering
\includegraphics[width=2.8in]{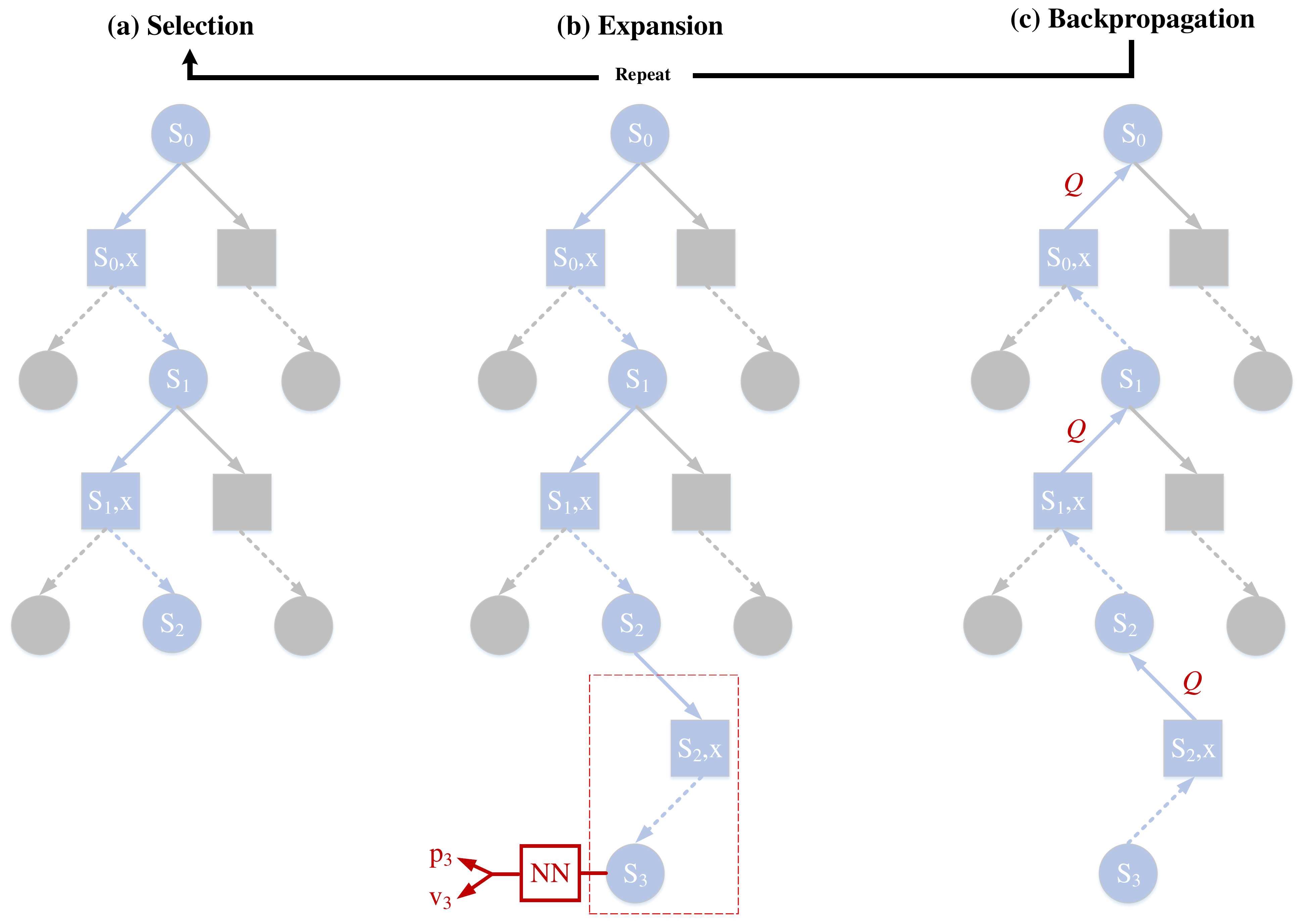}
\caption{MCTS in AlphaZero-UVR.} \label{fig:MCTS in AlphaZero-UVR}
\vspace{-1.5em}
\end{figure}

\textbf{Selection:} Each simulation always starts from the root node $S_0$, and finishes when it reaches a leaf node $S_L$.
For each hypothetical time-step $l=1, 2, \cdots, L$ of the selection stage, we select action $a_l$ according to the stored statistics for node $S_{l-1}$ using the PUCT strategy:
\begin{equation}\label{EQ11}
\begin{aligned}
a_l = \arg\max_{a} \bigg\{Q(S, a) + c_{puct}P(S, a) \frac{\sqrt{\sum_b N(S, b)}}{1+N(S, a)}\bigg\}
 \end{aligned}
\end{equation}
where, $c_{puct}$ is a constant value. 
After taking the selected action, it reaches a random node $(S_{l-1}, a_l)$.
Then it generates a damage observation $o$ of the passed line according to the current state information of the system, and it reaches a decision node.
This process repeat until the final time-step $L$.

\textbf{Expansion:} 
At the final time-step $L$ of the simulation, the state $S_L$ and reward  $R(S_{L-1}, a_L) = r_L(S_{L-1}, a_L)$ are respectively computed by the transition function and the reward function (\ref{EQ10}) according to $S_{L-1}$, $a_L$, and $o_L$.
A new random node, corresponding to state-action $(S_{L-1}, a_L)$ is added to the tree, and a new decision node, corresponding to state $S_L$ is also added to the tree as shown in Fig. \ref{fig:MCTS in AlphaZero-UVR}.
Each edge $(S_{L-1}, a, o)$ is initialized to:
\begin{equation}\label{EQ12}
\begin{aligned}
\bigg\{N(S_{L-1}, a_L, o) = 0\bigg\}
 \end{aligned}
\end{equation}
In the same time, the statistics $(p_a, v_L)$ of the new node $S_L$ is computed by the neural network:
\begin{equation}\label{EQ13}
\begin{aligned}
(p_a, v_L) = f_{\theta_{}}(S_L)
 \end{aligned}
\end{equation}
where, $\theta$ represents the parameters of the neural network.
This neural network takes system state as input and outputs action probabilities \textbf{p} with components $p_a = Pr(a|S)$ for each action $a$, and a scalar value $v$ which represents the estimation of the expected outcome $z$ from current state.
And each edge $(S_L, a)$ is initialized to:
\begin{equation}\label{EQ14}
\begin{aligned}
\bigg\{N(S_L, a) = 0, W(S_L, a) = 0, Q(S_L, a) = 0, P(S_L, a) = p_a\bigg\}
 \end{aligned}
\end{equation}
Note that the reward information of each edge $(S_L, a)$ is also initialized to zero and will be calculated using Eq. (\ref{EQ10}) when we visit the corresponding edge.

\textbf{Backpropagation:} At the end of the simulation, the statistics of the passed edges are updated back through the path.
For $l = L, \cdots, 1$, the statistics of the edges in the simulation paths are updated by:
\begin{equation}\label{EQ15}
\begin{aligned}
&N(S_{l-1}, a_l) = N(S_{l-1}, a_l) + 1 \\
&N(S_{l-1}, a_l, o_l) = N(S_{l-1}, a_l, o_l) + 1 \\
&W(S_{l-1}, a_l) = W(S_{l-1}, a_l) + G_l \\
&Q(S_{l-1}, a_l) = \frac{W(S_{l-1}, a_l)}{N(S_{l-1}, a_l)}
 \end{aligned}
\end{equation}
where $G_l$ is an $L-l$-step estimate of the cumulative discounted reward, bootstrapping from $v_L$:
\begin{equation}\label{EQ16}
\begin{aligned}
G^l = \sum_{\varsigma = 0}^{L-1-l} \gamma^{\varsigma} r_{l+1+\varsigma} + \gamma^{L-l} v_L
 \end{aligned}
\end{equation}
where $\gamma$ is the discount factor.
$r_{l+1+\varsigma}$ represents the reward function at the hypothetical time-step $l+1+\varsigma$.
$v_L$ is the value computed in Eq. (\ref{EQ13}).

At each time-step $t$, starting from the root position $S_t$ (used to generate the root node $S_0$), an MCTS search $\pi_t = \alpha_{\theta_{i-1}}(S_t)$ which consists of hundreds of simulations shown in (\ref{EQ11}) - (\ref{EQ16}) is executed using the previous iteration of neural network $f_{\theta_{i-1}}$.
After we finished the MCTS search, an action is sampled according to the obtained search probabilities $\pi_t$.
The search probabilities $\pi_t$ is actually a vector that represents the selection probability of each feasible action $a_t$ when system in current state $S_t$, and the probability is proportional to the exponentiated visit count for each action:
\begin{equation}\label{EQ17}
\begin{aligned}
\pi_{a_t} \propto N(S_t, a_t)^{\frac{1}{\tau}}
 \end{aligned}
\end{equation} 
where $\tau$ is a temperature parameter.

\textit{3) \textbf{Immediate reward signal}}: For board games, the agent does not receive the immediate reward signal at the intermediate time-steps.
When game terminates at step $T$, then the game is scored to a final reward $r_T \in \{-1, 0, 1\}$ according to the rules of win/loss/draw.
And the data of each time-step $t$, $(S_t, \pi_t, z_t)$, is stored in the replay buffer, where $z_t = \pm r_T$.
To this end, a game is played and the training data are generated by self-play.

However, for the UVR problem, there exists an immediate reward signal after the agent takes a decision, as shown in (\ref{EQ10}).
Thus, it needs to design the value target $z_t$ which corresponds to the cumulative reward function of the UVR problem.
In this paper, the authors use the result of the MCTS search as a target value estimator \cite{moerland2018a0c}, leveraging the action value estimates $Q(S_t, a)$ at the root state $S_t$.
The designed target value estimation method is given in (\ref{EQ18}).
\begin{equation}\label{EQ18}
\begin{aligned}
z_t = \max_{a} Q(S_t, a)
 \end{aligned}
\end{equation}
For each actual time-step $t$, the generated vehicle routing data $(S_t, \pi_t, z_t)$ is stored in the replay buffer and used to train the neural network.

\textit{4) \textbf{Value function normalization}}: The range of the value function $V$ of the game of Go, shogi, and chess is $[-1, 1]$, which is very helpful for the algorithm convergence.
However, the reward/value range of the UVR problem is $(-\inf, 0)$, as shown in Eq. (\ref{EQ10}).
The unbounded value range will bring difficulties to the training of the AlphaZero-UVR algorithm and even causes the algorithm to diverge \cite{schadd2008single}.
To deal with this problem, we normalize $Q$ value within $[0, 1]$ interval by using the minimum-maximum values observed in the search tree up to that point \cite{schrittwieser2019mastering}.
When a node is reached during the selection procedure, the normalized $\bar Q$ value is computed by:
\begin{equation}\label{EQ19}
\begin{aligned}
\bar Q(S_t, a) = \frac{Q(S_t, a)-\min_{S, a \in Tree} Q(S, a)}{\max_{S, a \in Tree} Q(S, a) - \min_{S, a \in Tree} Q(S, a)}
 \end{aligned}
\end{equation}
Thus, in the above equation, the normalized value $\bar Q(S_t, a)$ is utilized to replace the original value $Q(S_t, a)$ in Eq. (\ref{EQ11}).

Finally, during the self-play process, new network parameters $\theta_i$ are trained in parallel by uniformly sampling data from the replay buffer.
The new network $f_{\theta_i}(\cdot)$ is adjusted to minimize the following losses:
\begin{equation}\label{EQ20}
\begin{aligned}
l = (z-v)^2 - \pi ^\intercal log \textbf{p} + c ||\theta||^2
 \end{aligned}
\end{equation}
where, $\textbf{p}$ and $v$ are the policy and the value output of the new neural network $f_{\theta_i}(\cdot)$.
$\theta$ represents the weights of the neural network.
$\pi$ and $z$ are the sampled policy and value data from the reply buffer.
$c$ is a parameter to prevent overfitting.

\subsection{Implementation details of the AlphaZero-UVR algorithm}   
The AlphaZero-UVR algorithm is trained off-line first to get a well-trained neural network model.
Then, we apply the well-trained agent to navigate the post-storm restoration of the same distribution system sequentially according to the actual state of the system.
Using the well-trained model, the application process of the AlphaZero-UVR algorithm is shown in \textbf{Algorithm 1}.
At each time-step, the EUC checks if there are scheduling requests from vehicles. If it receives any request, the agent will get the current state information of the system as shown in equation (3), and construct the root node using current state information.
Then, starting from the root node, the agent conducts a predefined number of MCTS simulations to set up a search tree.
And each MCTS simulation consists of selection, expansion, and backpropagation.
Note that the well-trained neural network model will be used in expansion stage to calculate the action policy and value of the new node as shown in Eq. (\ref{EQ13}). 
After getting the search tree, the visiting times of each feasible routing actions of the root node can be easily obtained.
The action with the highest visiting times is selected as the optimal routing action.
Next, the vehicle travels to the destination node and gets the damage information of the power line it passed through.
If the passed power line is damaged, the vehicle will repair it.
The vehicle also needs to report the actual damage status of the passed line to the EUC.
Finally, if current scheduling request queue is empty, the agent calculates the updated belief state of the distribution system using equation (1).
The repairing process stops when the damage probability of each line is below a threshold $\epsilon ^{thr}$.
It can be found that the optimal repair trajectory of the vehicles are calculated online according to the real-time status of the distribution system.
The neural network model used in the MCTS simulations needs to be trained off-line before the on-line application.
The details of the training process of the AlphaZero-UVR algorithm are provided in the Supplementary Materials (see \textbf{Algorithm S1}).
\begin{algorithm}   \caption{AlphaZero based Utility Vehicle Routing Approach.}\label{alg:AlphaZero-UVR}
\small
\begin{algorithmic}[1]
    \State Initialize the state $S_0$. Set time-step $t = 0$. 
	\While {$p(L_{t, i}^e = 1|H_t, A_{t-1}) \geq \epsilon ^{thr}$}: \Comment {\textit{{\footnotesize Stop repairing the grid until the damage probability of each line is below a threshold.}}}
		\begin{enumerate}
			\item At time-step $t$, get the physical state (according to the scheduling requests from vehicles), belief state, and informational state of the distribution system defined in Eq. (\ref{EQ3}).
			\item Use MCTS with the learned neural network to solve Eq. (\ref{EQ9}) that determines the optimal action. \Comment {\textit{{\footnotesize Perform a number of MCTS simulations to construct the search tree and the action with the highest visiting times is the optimal action.}}}
			\item Move the utility vehicle, which sends the scheduling request to the EUC, according to the obtained optimal action, then get the observation of the damage status of the passed line.
			\item Delete the dispatched vehicle from current scheduling request queue. If the scheduling request queue is not empty, select the next vehicle from the queue and go back to step 1).
			\item Update the belief state of the distribution system using Eq. (\ref{EQ1}).
			\item $t = t +1$.
		\end{enumerate}
	\EndWhile 
\end{algorithmic}
\end{algorithm}

\vspace{-1em}
\section{Simulation Results}  \label{Simulation Results}
In this section, the effectiveness of the AlphaZero-UVR algorithm is demonstrated by the numerical simulations on two distribution systems.
All the simulations were conducted on an Intel Core i7 @1.90 GHz Windows-based PC with 16GB RAM.
We implemented the proposed AlphaZero-UVR algorithm using Tensorflow library in Python.

\subsection{Case study I: 8-node test system}
The first test distribution system used in the simulation is shown in Fig. \ref{fig:DistributionPowerSystem}.
In this case study, we focus on the single utility vehicle routing problem which is a special case of multi-vehicle routing.
It is assumed that the vehicle always depots from the substation.
The prior damage probabilities of the power lines after a storm are given in Fig. \ref{fig:DistributionPowerSystem}, which can be estimated by storm weather information and operator's experience.
In addition, the number of customers and the time for the vehicle passing through each road are also shown in the figure.
The calling probability of customers when suffering from outages is set to $\rho = 5\%$.
The stopping threshold $\epsilon^{thr}$ is set to $2\%$.
To simplify the problem, the required repair time for each line is assumed to be 1 hour.

At each time-step $t$, the state $(G_t, P_t^{L, post})$ is fed to the neural network of the agent to get the policy and value.
So, for the restoration problem shown in Fig. \ref{fig:DistributionPowerSystem}, the number of neurons of the input layer is 8.
Followed the input layer is two hidden layers and each hidden layer is with 120 neurons.
The output layer contains 5 neurons.
Note that the maximum feasible actions of the vehicle in Fig. \ref{fig:DistributionPowerSystem} is 4, so the action probabilities output by the neural network is a 4 dimensional vector.
Besides, the learning rate of the algorithm is set to 0.0001 and the batch size $B=32$.
The RMSprop optimizer is adopted to train the neural network. 

\begin{figure}[t]\centering
\includegraphics[width=3.0in]{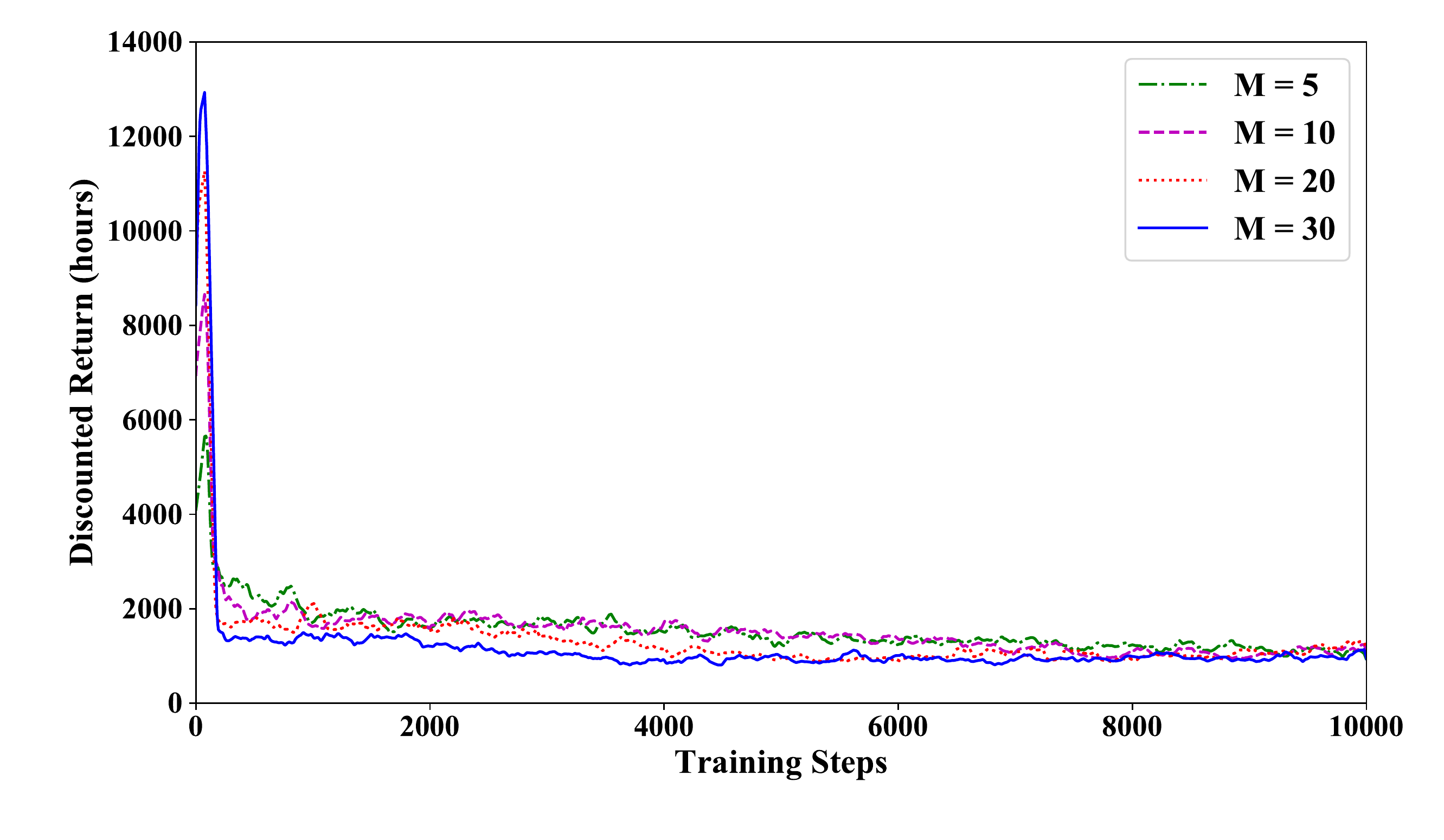}
\caption{The convergence process of the AlphaZero-UVR algorithm under different number of simulations per move.} \label{fig:Convergence1}
\vspace{-0.5em}
\end{figure}

The number of simulations ($M$) of the MCTS search procedure in Algorithm S1 affects the performance of the proposed algorithm.
To analyze the sensitivity of the performance of the algorithm with respect to the number of simulations per move, we tested the convergence performance of the proposed AlphaZero-UVR algorithm under different $M$ value, as shown in Fig. \ref{fig:Convergence1}.
It can be found that the cumulative outage hours of the customers decreases rapidly in the first 500 training steps, and then it slowly approaches to the optimal value.
Besides, with more simulations per move, the algorithm can achieve better optimization performance.

To validate the effectiveness of the proposed algorithm, we compared the performance of the algorithm with traditional MCTS \cite{al2016information} and the OLUCT method \cite{hang2020IEEEPES}.
The optimization performance of the traditional tree search methods adopted in \cite{al2016information} and \cite{hang2020IEEEPES} are also influenced by the number of simulations per move.
For the comparing methods, we set $M$ to 200 using which both methods perform very well, while the $M$ value of the proposed algorithm is set to 30.
The simulation results are shown in Table I.
In the table, the performance of the algorithms were tested using 10 different fault settings.
For each case setting, the actual damaged lines are preset, which is unknown to the agent, and the phone calls (received or not) from customer node $[2, 4, 5, 6, 7]$ are also given.

From the above results, it can be found that the proposed algorithm outperforms the other two MCTS based methods even if the number of simulations per move of the proposed algorithm is much less than the other methods.
The good performance of the proposed algorithm can be attributed to the guidance of the well-trained neural network model during the tree search process.
Besides, the proposed algorithm has better computational efficiency.
The computational time required for a single time step scheduling for the proposed algorithm is 0.76s, and the corresponding computational time of the traditional MCTS and the OLUCT methods are 6.7s and 19s, respectively.
\renewcommand{\arraystretch}{1.25} 
\begin{table*}[t]
\scriptsize
\centering
\caption{The optimized customer outage hour (h) using different algorithms for the 8-node test system.}
\label{Tab03}
\begin{tabular}{cccccccccc}
\toprule
\multirow{2}{*}{Case} & \multicolumn{2}{c}{Case setting} & \multicolumn{2}{c}{AlphaZero-UVR} & \multirow{2}{*}{$F_{MCTS}$ (hours)} & \multirow{2}{*}{$F_{OLUCT}$ (hours)} \\
\cmidrule(r){2-3} \cmidrule(r){4-5}
&  Phone calls      &  Damaged lines
&  $F_{AlphaZero-UVR}$ (hours)      &  Repairing trajectory  \\
\midrule
$1$             &[0 0 1 0 0]                          & [$L_5$]                    & 250                   & 0 $\longrightarrow$ 1 $\longrightarrow$ 2 $\longrightarrow$ 3 $\longrightarrow$ 5           & 250 & 250 \\
$2 $             &[0 0 0 0 1]                          & [$L_7$]                    & 383.3                  & 0 $\longrightarrow$ 1 $\longrightarrow$ 6 $\longrightarrow$ 1 $\longrightarrow$ 2 $\longrightarrow$ 6 $\longrightarrow$ 7          &516.67           & 516.67                   \\
3             &[0 1 1 0 1]                         &[$L_3$ $L_7$]                   &1016.7 &\tabincell{c}{0 $\longrightarrow$ 1 $\longrightarrow$ 2 $\longrightarrow$ 3 $\longrightarrow$ 5 $\longrightarrow$ 3  \\ $\longrightarrow$ 2 $\longrightarrow$ 4 $\longrightarrow$ 2 $\longrightarrow$ 6 $\longrightarrow$ 7} &1516.7           &1216.7                     \\
$4 $             &[1 1 1 0 0]                         &[$L_2$ $L_3$]                   &850                   & 0 $\longrightarrow$ 1 $\longrightarrow$ 2 $\longrightarrow$ 4 $\longrightarrow$ 2 $\longrightarrow$ 3 $\longrightarrow$ 5          &716.66          & 716.66                  \\
$5$   &[1 1 1 0 0]                         &[$L_2$ $L_4$]                   &683.3                   & 0 $\longrightarrow$ 1 $\longrightarrow$ 2 $\longrightarrow$ 4 $\longrightarrow$ 2 $\longrightarrow$ 3 $\longrightarrow$ 5    &1016.7          & 850.0
\\
$6$   &[0 1 1 0 0]                         &[$L_3$ $L_4$]                   &766.7                   & 0 $\longrightarrow$ 1 $\longrightarrow$ 2 $\longrightarrow$ 4 $\longrightarrow$ 2 $\longrightarrow$ 3 $\longrightarrow$ 5    &766.7          & 766.7
\\
$7 $             &[0 1 1 0 0]                         &[$L_4$ $L_5$]                  &616.7                   & 0 $\longrightarrow$ 1 $\longrightarrow$ 2 $\longrightarrow$ 4 $\longrightarrow$ 2 $\longrightarrow$ 3 $\longrightarrow$ 5          &1066.67          & 1066.67
\\
$8$   &[0 0 1 0 1]                         &[$L_5$ $L_7$]                   &733.3                   & 0 $\longrightarrow$ 1 $\longrightarrow$ 6 $\longrightarrow$ 7 $\longrightarrow$ 6 $\longrightarrow$ 1   &1266.7          & 1200.0
\\
$9$             &[0 1 1 0 0]                         &[$L_3$ $L_4$ $L_5$]                   &900                   & 0 $\longrightarrow$ 1 $\longrightarrow$ 2 $\longrightarrow$ 4 $\longrightarrow$ 2 $\longrightarrow$ 6 $\longrightarrow$ 7          &1133          & 1133
\\
$10$   &[0 1 1 0 1]                         &[$L_3$ $L_4$ $L_7$]                   &1516.7                   & \tabincell{c}{0 $\longrightarrow$ 1 $\longrightarrow$ 2 $\longrightarrow$ 4 $\longrightarrow$ 2 $\longrightarrow$ 3 $\longrightarrow$ 5 \\ $\longrightarrow$ 3 $\longrightarrow$ 5 $\longrightarrow$ 3 $\longrightarrow$ 2 $\longrightarrow$ 6 $\longrightarrow$ 7}    &1616.7          & 1616.7
\\
\bottomrule
\end{tabular}
\vspace{-2.0em}
\end{table*}
\vspace{-1.0em}

\subsection{Case study II: modified IEEE 123-node test system}
\begin{figure}[t]\centering
\includegraphics[width=3.2in]{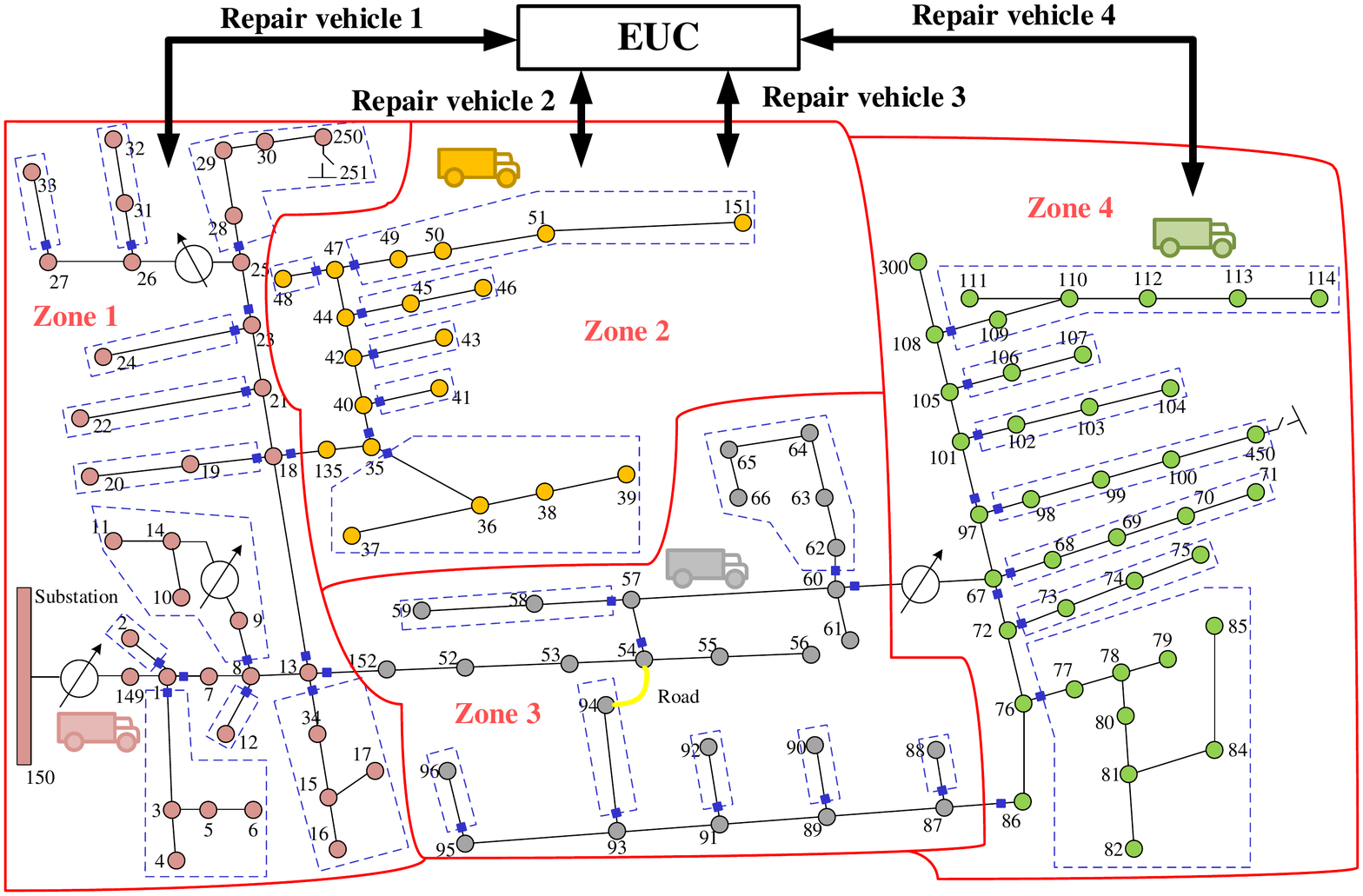}
\caption{The modified IEEE 123-node distribution system. The blue rectangles represent the locations of protective devices, and the dashed frame indicates the protection scope of the protective devices. The red line represents boundaries of each zone.} \label{fig:IEEE 123 bus system topology}
\vspace{-1.5em}
\end{figure}

To further validate the effectiveness of the proposed algorithm, we tested the performance of the algorithm on a modified IEEE 123-node distribution system, as shown in Fig. \ref{fig:IEEE 123 bus system topology}.
It is assumed that the road paths are along each lines in the system.
The locations of protective devices are placed according to reference \cite{butler2009overcurrent}, as shown in Fig. \ref{fig:IEEE 123 bus system topology}.
The distribution system contains 123 nodes and 197 lines.
The parameters of the modified IEEE 123-node system can be found in \cite{IEEE123}.
Reference \cite{IEEE123} also provided the length of each power line.
In this work, we enlarged the length of each line to expand the coverage area of the distribution grid, and the average travel velocity of each vehicle is assumed to be 20 miles per hour.
According to the topology of the test system, we divide the system into four zones, and each vehicle is responsible for one specific zone. 
The repair time for a damaged line in zone 1 and zone 3 is assumed to be 60 minutes, and the repair time for a damaged line in other zones is 120 minutes \cite{9115714}.
In addition, the vehicle 1 has the highest priority, and the vehicle 4 has the lowest priority.
To decrease the computational complexity of the fault probability model shown in equation (1), we aggregated the power lines of the same segment and adopted Monte Carlo simulation to obtain the approximated fault probabilities of the segments.
After aggregation, the system contains 62 segments and 42 customer nodes.

The designed neural network model is a four-layer fully connected network.
As the state information consists of the position information of the vehicle that should be scheduled immediately and the fault probability of each segment, the number of neurons of the input layer is 63.
There are two hidden layers and each layer contains 150 neurons.
Note that the maximum feasible actions of the vehicles in this case study is 3, so the action policy output by the neural network is a three-dimensional vector.
Estimated value is also output by the neural network, so the output layer contains 4 neurons.
The learning rate of the algorithm is set to 0.001, and all the other hyperparameters are the same with case study I.

\begin{figure}[t]\centering
\includegraphics[width=2.8in]{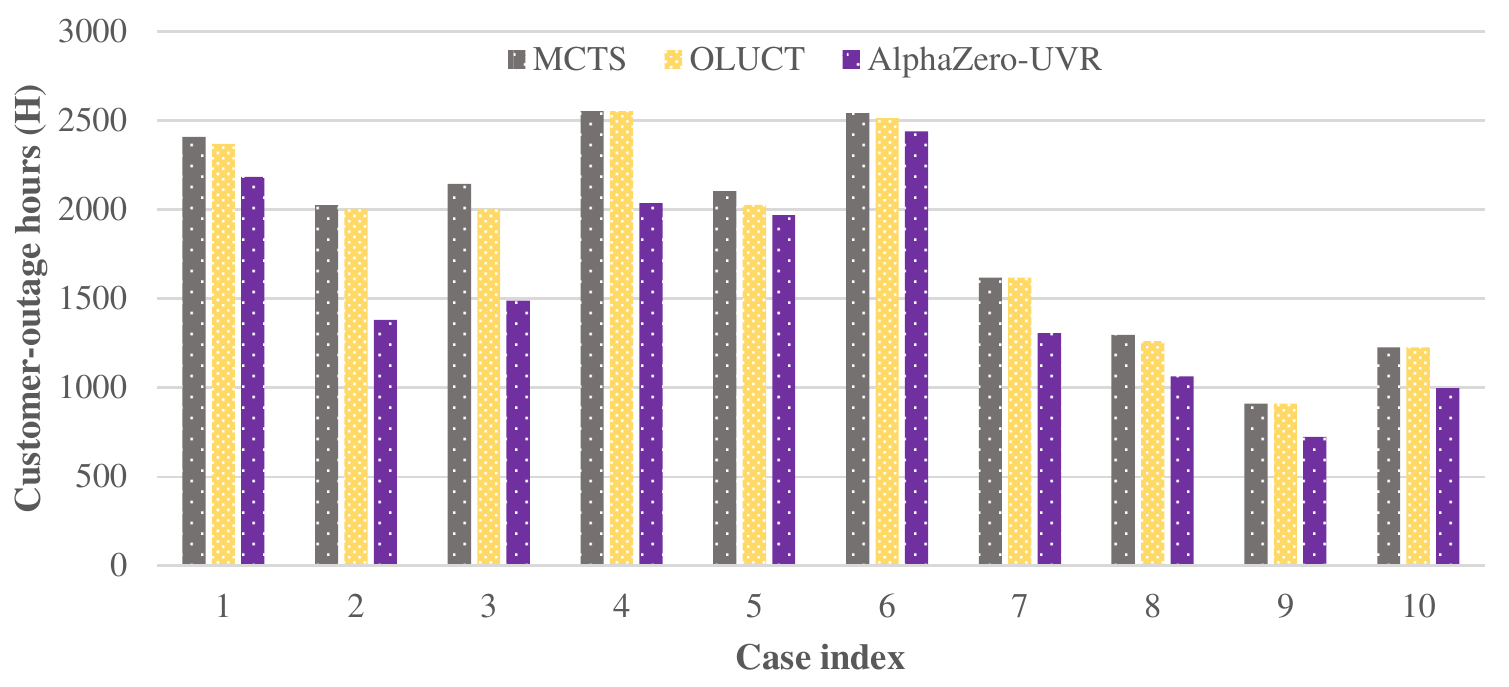}
\caption{The performance of the proposed algorithm and the comparing methods on the modified IEEE 123-node test system.} \label{fig:IEEE123Result}
\vspace{-1.2em}
\end{figure}

We also compared the performance of the proposed algorithm with the traditional MCTS and the OLUCT method.
The methods were compared under ten different cases, and each case was with different damage sets.
The results are shown in Fig. \ref{fig:IEEE123Result}.
It can be found that the proposed method performs better than the comparing methods.
The results demonstrated the effectiveness of the proposed multi-vehicle routing algorithm.

\vspace{-1em}
\section{Conclusion} \label{Conclusion}
In this work, an AlphaZero based post storm utility vehicle routing algorithm was proposed to guide repair crews in multiple vehicles to fix the damages in the distribution grid as fast as possible.
The utility vehicle routing optimization problem was modeled as a Markov Decision Process (MDP).
Then, the challenges of using the AlphaZero algorithm to solve the optimization problem in this work were presented and the corresponding solutions were proposed.
The proposed AlphaZero-UVR algorithm combined stochastic Monte-Carlo tree search with deep neural network, and can teach itself to navigate the crews to restore the distribution grid through self-play.
To validate the effectiveness of the proposed algorithm, the performance of the algorithm was tested by numerical simulations on a 8-node distribution grid and a modified IEEE 123-node distribution grid.
Simulation results demonstrated that the proposed algorithm outperforms traditional MCTS based methods.

\bibliographystyle{IEEEtran}
\bibliography{AlphaZero-UVR}

\begin{thebibliography}{10}
\providecommand{\url}[1]{#1}
\csname url@samestyle\endcsname
\providecommand{\newblock}{\relax}
\providecommand{\bibinfo}[2]{#2}
\providecommand{\BIBentrySTDinterwordspacing}{\spaceskip=0pt\relax}
\providecommand{\BIBentryALTinterwordstretchfactor}{4}
\providecommand{\BIBentryALTinterwordspacing}{\spaceskip=\fontdimen2\font plus
\BIBentryALTinterwordstretchfactor\fontdimen3\font minus
  \fontdimen4\font\relax}
\providecommand{\BIBforeignlanguage}[2]{{%
\expandafter\ifx\csname l@#1\endcsname\relax
\typeout{** WARNING: IEEEtran.bst: No hyphenation pattern has been}%
\typeout{** loaded for the language `#1'. Using the pattern for}%
\typeout{** the default language instead.}%
\else
\language=\csname l@#1\endcsname
\fi
#2}}
\providecommand{\BIBdecl}{\relax}
\BIBdecl

\bibitem{silver2017mastering}
D.~Silver, J.~Schrittwieser, K.~Simonyan, I.~Antonoglou, A.~Huang, A.~Guez,
  T.~Hubert, L.~Baker, M.~Lai, A.~Bolton \emph{et~al.}, ``Mastering the game of
  go without human knowledge,'' \emph{Nature}, vol. 550, no. 7676, pp.
  354--359, 2017.

\bibitem{kenward2014blackout}
A.~Kenward and U.~Raja, ``Blackout: Extreme weather climate change and power
  outages,'' \emph{Climate central}, vol.~10, pp. 1--23, 2014.

\bibitem{house2013presidential}
W.~House, ``Presidential policy directive--critical infrastructure security and
  resilience,'' \emph{Press Release, February}, vol.~12, 2013.

\bibitem{guikema2014predicting}
S.~D. Guikema, R.~Nateghi, S.~M. Quiring, A.~Staid, A.~C. Reilly, and M.~Gao,
  ``Predicting hurricane power outages to support storm response planning,''
  \emph{IEEE Access}, vol.~2, pp. 1364--1373, 2014.

\bibitem{yuan2016robust}
W.~Yuan, J.~Wang, F.~Qiu, C.~Chen, C.~Kang, and B.~Zeng, ``Robust
  optimization-based resilient distribution network planning against natural
  disasters,'' \emph{IEEE Trans. Smart Grid}, vol.~7, no.~6, pp. 2817--2826,
  2016.

\bibitem{amirioun2017resilience}
M.~Amirioun, F.~Aminifar, and H.~Lesani, ``Resilience-oriented proactive
  management of microgrids against windstorms,'' \emph{IEEE Trans. Power
  Syst.}, vol.~33, no.~4, pp. 4275--4284, 2017.

\bibitem{qiu2015optimal}
F.~Qiu, J.~Wang, C.~Chen, and J.~Tong, ``Optimal black start resource
  allocation,'' \emph{IEEE Trans. Power Syst.}, vol.~31, no.~3, pp. 2493--2494,
  2015.

\bibitem{yan2018coordinated}
M.~Yan, Y.~He, M.~Shahidehpour, X.~Ai, Z.~Li, and J.~Wen, ``Coordinated
  regional-district operation of integrated energy systems for resilience
  enhancement in natural disasters,'' \emph{IEEE Trans. Smart Grid}, vol.~10,
  no.~5, pp. 4881--4892, 2018.

\bibitem{trakas2017optimal}
D.~N. Trakas and N.~D. Hatziargyriou, ``Optimal distribution system operation
  for enhancing resilience against wildfires,'' \emph{IEEE Trans. Power Syst.},
  vol.~33, no.~2, pp. 2260--2271, 2017.

\bibitem{hou2011computation}
Y.~Hou, C.-C. Liu, K.~Sun, P.~Zhang, S.~Liu, and D.~Mizumura, ``Computation of
  milestones for decision support during system restoration,'' in \emph{2011
  IEEE Power and Energy Society General Meeting}.\hskip 1em plus 0.5em minus
  0.4em\relax IEEE, 2011, pp. 1--10.

\bibitem{sun2010optimal}
W.~Sun, C.-C. Liu, and L.~Zhang, ``Optimal generator start-up strategy for bulk
  power system restoration,'' \emph{IEEE Trans. Power Syst.}, vol.~26, no.~3,
  pp. 1357--1366, 2010.

\bibitem{arif2018optimizing}
A.~Arif, S.~Ma, Z.~Wang, J.~Wang, S.~M. Ryan, and C.~Chen, ``Optimizing service
  restoration in distribution systems with uncertain repair time and demand,''
  \emph{IEEE Trans. Power Syst.}, vol.~33, no.~6, pp. 6828--6838, 2018.

\bibitem{van2011vehicle}
P.~Van~Hentenryck, C.~Coffrin, R.~Bent \emph{et~al.}, ``Vehicle routing for the
  last mile of power system restoration,'' in \emph{Proceedings of the 17th
  Power Systems Computation Conference (PSCC 11), Stockholm, Sweden}.\hskip 1em
  plus 0.5em minus 0.4em\relax Citeseer, 2011.

\bibitem{7812566}
A.~{Arif}, Z.~{Wang}, J.~{Wang}, and C.~{Chen}, ``Power distribution system
  outage management with co-optimization of repairs, reconfiguration, and dg
  dispatch,'' \emph{IEEE Transactions on Smart Grid}, vol.~9, no.~5, pp.
  4109--4118, 2018.

\bibitem{9115714}
T.~{Ding}, Z.~{Wang}, W.~{Jia}, B.~{Chen}, C.~{Chen}, and M.~{Shahidehpour},
  ``Multiperiod distribution system restoration with routing repair crews,
  mobile electric vehicles, and soft-open-point networked microgrids,''
  \emph{IEEE Transactions on Smart Grid}, vol.~11, no.~6, pp. 4795--4808, 2020.

\bibitem{8642442}
S.~{Lei}, C.~{Chen}, Y.~{Li}, and Y.~{Hou}, ``Resilient disaster recovery
  logistics of distribution systems: Co-optimize service restoration with
  repair crew and mobile power source dispatch,'' \emph{IEEE Transactions on
  Smart Grid}, vol.~10, no.~6, pp. 6187--6202, 2019.

\bibitem{chen2017modernizing}
C.~Chen, J.~Wang, and D.~Ton, ``Modernizing distribution system restoration to
  achieve grid resiliency against extreme weather events: an integrated
  solution,'' \emph{Proc. IEEE}, vol. 105, no.~7, pp. 1267--1288, 2017.

\bibitem{7444207}
H.~{Sun}, Z.~{Wang}, J.~{Wang}, Z.~{Huang}, N.~{Carrington}, and J.~{Liao},
  ``Data-driven power outage detection by social sensors,'' \emph{IEEE
  Transactions on Smart Grid}, vol.~7, no.~5, pp. 2516--2524, 2016.

\bibitem{1007917}
{Yan Liu} and N.~N. {Schulz}, ``Knowledge-based system for distribution system
  outage locating using comprehensive information,'' \emph{IEEE Transactions on
  Power Systems}, vol.~17, no.~2, pp. 451--456, 2002.

\bibitem{956755}
K.~{Sridharan} and N.~N. {Schulz}, ``Outage management through amr systems
  using an intelligent data filter,'' \emph{IEEE Transactions on Power
  Delivery}, vol.~16, no.~4, pp. 669--675, 2001.

\bibitem{executive}
U.~D. of~Energy, ``Economic benefits of increasing electric grid resilience to
  weather outages,'' \emph{Executive Office of the President}, Aug. 2013.

\bibitem{bahmanyar2017comparison}
A.~Bahmanyar, S.~Jamali, A.~Estebsari, and E.~Bompard, ``A comparison framework
  for distribution system outage and fault location methods,'' \emph{Electric
  Power Systems Research}, vol. 145, pp. 19--34, 2017.

\bibitem{al2016information}
L.~Al-Kanj, W.~B. Powell, and B.~Bouzaiene-Ayari, ``The information-collecting
  vehicle routing problem\: Stochastic optimization for emergency storm
  response,'' \emph{arXiv preprint arXiv:1605.05711}, 2016.

\bibitem{hang2020IEEEPES}
H.~Shuai, H.~He, and J.~Wen, ``Post-storm vehicle routing for distribution grid
  restoration: An oluct based learning approach,'' in \emph{2020 IEEE Power and
  Energy Society General Meeting}.\hskip 1em plus 0.5em minus 0.4em\relax IEEE,
  2020, pp. 1--5.

\bibitem{silver2016mastering}
D.~Silver, A.~Huang, C.~J. Maddison, A.~Guez, L.~Sifre, G.~Van Den~Driessche,
  J.~Schrittwieser, I.~Antonoglou, V.~Panneershelvam, M.~Lanctot \emph{et~al.},
  ``Mastering the game of go with deep neural networks and tree search,''
  \emph{Nature}, vol. 529, no. 7587, p. 484, 2016.

\bibitem{FranLi2018}
F.~Li and Y.~Du, ``From alphago to power system ai: What engineers can learn
  from solving the most complex board game,'' \emph{IEEE Power and Energy
  Magazine}, vol.~16, no.~2, pp. 76--84, 2018.

\bibitem{schrittwieser2020mastering}
J.~Schrittwieser, I.~Antonoglou, T.~Hubert, K.~Simonyan, L.~Sifre, S.~Schmitt,
  A.~Guez, E.~Lockhart, D.~Hassabis, T.~Graepel \emph{et~al.}, ``Mastering
  atari, go, chess and shogi by planning with a learned model,'' \emph{Nature},
  vol. 588, no. 7839, pp. 604--609, 2020.

\bibitem{Hang2021TSG}
H.~Shuai and H.~He, ``Online scheduling of a residential microgrid via
  monte-carlo tree search and a learned model,'' \emph{IEEE Transactions on
  Smart Grid}, vol.~12, no.~2, pp. 1073--1087, 2021.

\bibitem{al2015probability}
L.~Al-Kanj, B.~Bouzaiene-Ayari, and W.~B. Powell, ``A probability model for
  grid faults using incomplete information,'' \emph{IEEE Trans. Smart Grid},
  vol.~8, no.~2, pp. 956--968, 2015.

\bibitem{moerland2018a0c}
T.~M. Moerland, J.~Broekens, A.~Plaat, and C.~M. Jonker, ``A0c: Alpha zero in
  continuous action space,'' \emph{arXiv preprint arXiv:1805.09613}, 2018.

\bibitem{schadd2008single}
M.~P. Schadd, M.~H. Winands, H.~J. Van Den~Herik, G.~M.-B. Chaslot, and J.~W.
  Uiterwijk, ``Single-player monte-carlo tree search,'' in \emph{International
  Conference on Computers and Games}.\hskip 1em plus 0.5em minus 0.4em\relax
  Springer, 2008, pp. 1--12.

\bibitem{schrittwieser2019mastering}
J.~Schrittwieser, I.~Antonoglou, T.~Hubert, K.~Simonyan, L.~Sifre, S.~Schmitt,
  A.~Guez, E.~Lockhart, D.~Hassabis, T.~Graepel \emph{et~al.}, ``Mastering
  atari, go, chess and shogi by planning with a learned model,'' \emph{arXiv
  preprint arXiv:1911.08265}, 2019.

\bibitem{butler2009overcurrent}
K.~L. Butler-Purry and H.~B. Funmilayo, ``Overcurrent protection issues for
  radial distribution systems with distributed generators,'' in \emph{2009 IEEE
  Power \& Energy Society General Meeting}.\hskip 1em plus 0.5em minus
  0.4em\relax IEEE, 2009, pp. 1--5.

\bibitem{IEEE123}
\BIBentryALTinterwordspacing
``Ieee 123 node test feeder,'' (Date last accessed 5-Febrary-2021). [Online].
  Available: \url{https://site.ieee.org/pes-testfeeders/resources/}
\BIBentrySTDinterwordspacing

\end{thebibliography}

\end{document}